\documentclass[11pt,a4paper]{article}
\usepackage[utf8]{inputenc}
\usepackage{amsmath}
\usepackage{amsfonts}
\usepackage{amssymb}
\usepackage{graphicx}
\usepackage[left=2.5cm,right=2.5cm,top=3cm,bottom=3cm]{geometry}
\begin{document}

\newcommand{\qcd}{{\footnotesize QCD }}
\newcommand{\qgp}{{\footnotesize QGP }}
\newcommand{\mds}{{\footnotesize MDs }}
\newcommand{\mdss}{{\footnotesize MDs}}
\newcommand{\urq}{{\footnotesize URQMD }}
\newcommand{\urqq}{{\footnotesize URQMD}}
\newcommand{\ampt}{{\footnotesize AMPT }}
\newcommand{\amptt}{{\footnotesize AMPT}}
\newcommand{\mc}{{\footnotesize MC }}
\newcommand{\fls}{{fluctuations }}
\newcommand{\fl}{{fluctuations}}

\begin{center}
\begin{Large}\textbf{
Exploring Event-by-Event $\it{p}_{\rm T}$ Fluctuations in pp Collisions at $\sqrt{s} = 13$ TeV: An Insight from ALICE
}
\end{Large}\\ [1cm]

\textsf{Bushra Ali\footnote{Bushra.Ali@cern.ch}, for the ALICE Collaboration}\\
\textsf{Department of Physics, Aligarh Muslim University, India}\\ [1cm]
\end{center}

\noindent \textbf{Abstract}: Event-by-event fluctuations of the mean transverse momentum ($p_{\rm T}$) of relativistic charged particles are analyzed using the two-particle correlator $\sqrt{C_m}/M(p_{\rm T})_m$, which quantifies the correlations strength in units of the mean $p_{\rm T}$ in proton-proton collision at $\sqrt{s} = 13$ TeV in ALICE both for minimum bias and and high-multiplicity triggered events. The non-monotonic variations in $p_{\rm T}$ correlations with changing energy could serve as a signature of QGP formation. A comprehensive investigation across soft-, intermediate-, and hard-$p_{\rm T}$ regions could provide crucial insights into both equilibrium (e.g., thermal radial flow) and non-equilibrium (e.g., jet/minijet) contributions to $p_{\rm T}$ fluctuations. The dependence of the correlator on particle multiplicity for different $p_{\rm T}$ window widths and positions is explored. The correlator values are found to decrease with increasing charged particle density, following a power-law behavior similar to observations in both small and large systems at lower energies. Additionally, the influence of $p_{\rm T}$ range on the power-law coefficient is studied and results are compared with predictions from Monte Carlo models, such as PYTHIA (pQCD string model) and EPOS (core-corona model), to enhance understanding of the underlying mechanisms driving $p_{\rm T}$ fluctuations.\\ \\

\noindent \textsf{\bf The 10th Asian Triangle Heavy-Ion Conference - ATHIC 2025}\\
\noindent \textsf{\bf 13-16 Jan, 2025}\\
\noindent \textsf{\bf Indian Institute of Science Education and Research Berhampur, Odisha, India} \\ \\

\section{Introduction}
\indent Fluctuations in mean transverse momentum may emerge from various correlation sources, for example, including resonance decays, quantum statistical effects, and hard scatterings such as jet production. These contributions are not exclusive to nucleus--nucleus (AA) collisions rather they have also been observed in smaller systems, like proton--proton (pp) at LHC energies. Results from pp collisions, therefore, would serve as a model-independent baseline to look for the occurrence of fluctuation patterns of dynamic nature in AA collisions, which lead to the modification of the fluctuation patterns exhibited by pp collisions~\cite{5}. Moreover, recent observations of strangeness enhancement comparable to that observed in AA collisions, flow-like signatures, and presence of collectivity in high-multiplicity pp collision events have reignited the interest towards the investigations involving small systems, like pp, at LHC energies~\cite{14,15,16}. A dedicated analysis of mean $\it{p}_{\rm T}$ correlations in high-multiplicity pp events revealed that these events exhibit a similar trend of decreasing fluctuations with increasing multiplicity, consistent with the behavior observed in minimum bias data~\cite{17}.\\

\indent To probe the physical origin of $\langle p_{\rm T} \rangle$ fluctuations in detail, a differential analysis of $p_{\rm T}$ correlations has been performed across various $p_{\rm T}$ intervals. This approach is motivated by the fact that the soft and hard processes dominate different regions of the $p_{\rm T}$ spectrum: hydrodynamic collectivity is expected to be significant at low $p_{\rm T}$, while high $p_{\rm T}$ particles are more likely to originate from hard scatterings and parton fragmentation~\cite{16a, 16b}. Earlier studies involving experiments, such as STAR~\cite{9, 18, 19, 20, 20a}, ALICE~\cite{5}, CERES~\cite{21, 22}, and PHENIX~\cite{23} were primarily restricted to a limited $p_{\rm T}$ acceptance (0.15 $< p_{\rm T} <$ 2.0 GeV/$c$). Attempts are, therefore, made to study the mean $p_{\rm T}$ correlations in $p_{\rm T}$ windows having higher ranges, like 0.15 $< p_{\rm T} <$ 3.0, 0.15 $< p_{\rm T} <$ 4.0, and 0.15 $< p_{\rm T} <$ 6.0 GeV/$c$, to investigate how the inclusion of harder particles influences correlation strength. \\

\indent In order to isolate the $p_{\rm T}$ dependent effects, $p_{\rm T}$ windows of fixed widths ($1$ GeV/$c$ range) have also been considered: 1.0 $< p_{\rm T} <$ 2.0, 2.0 $< p_{\rm T} <$ 3.0, 3.0 $< p_{\rm T} <$ 4.0, and 4.0 $< p_{\rm T} <$ 5.0 GeV/$c$. In $p_{\rm T}$ windows positioned in regions of smaller $p_{\rm T}$, domination of soft processes is expected, whereas in $p_{\rm T}$ window placed in high $p_{\rm T}$ regions, hard scattering dynamics would dominate. A comparison of the experimental findings with the predictions of theoretical models, like PYTHIA8 (based on perturbative QCD and multiple parton interactions) and EPOS LHC (incorporating a core-corona framework with collective and jet-like features) would help understand the origins of mean $p_{\rm T}$ correlations. \\

\indent The two-particle $p_{\rm T}$ correlator $C$, in a multiplicity class m, is defined as~\cite{5},\\[-3mm]
\begin{equation}
C_m = \frac{1}{\Sigma_{k}^{n_{ev},m} N_k^{pairs}} \Sigma_{k=1}^{n_{ev},m}\Sigma_{i=1}^{N_{acc},k}\Sigma_{j=i+1}^{N_{acc},k} \left(\left(p_{\rm T, i} - M(p_{\rm T})_m\right) . \left(p_{\rm T, j} - M(p_{\rm T})_m\right)  \right),
\end{equation}
where $N_k^{pairs}$ is the number of pairs in an event, $n_{ev}$ is the number of events in class $m$ while $M(p_{\rm T})_m$ denotes the mean $p_{\rm T}$ of all particles in all events of that multiplicity class and may be estimated using the following relation:\\[-3mm]
\begin{equation}
M(p_{\rm T})_m = \frac{1}{\Sigma_{k=1}^{n_{ev},m} N_{acc, k}} \Sigma_{k=1}^{n_{ev},m}\Sigma_{i=1}^{N_{acc},k} p_{\rm T,i}.
\end{equation}
\section{Result and discussion}\label{}
\begin{figure}[htbp!]
  \centering
    \includegraphics[scale=0.36]{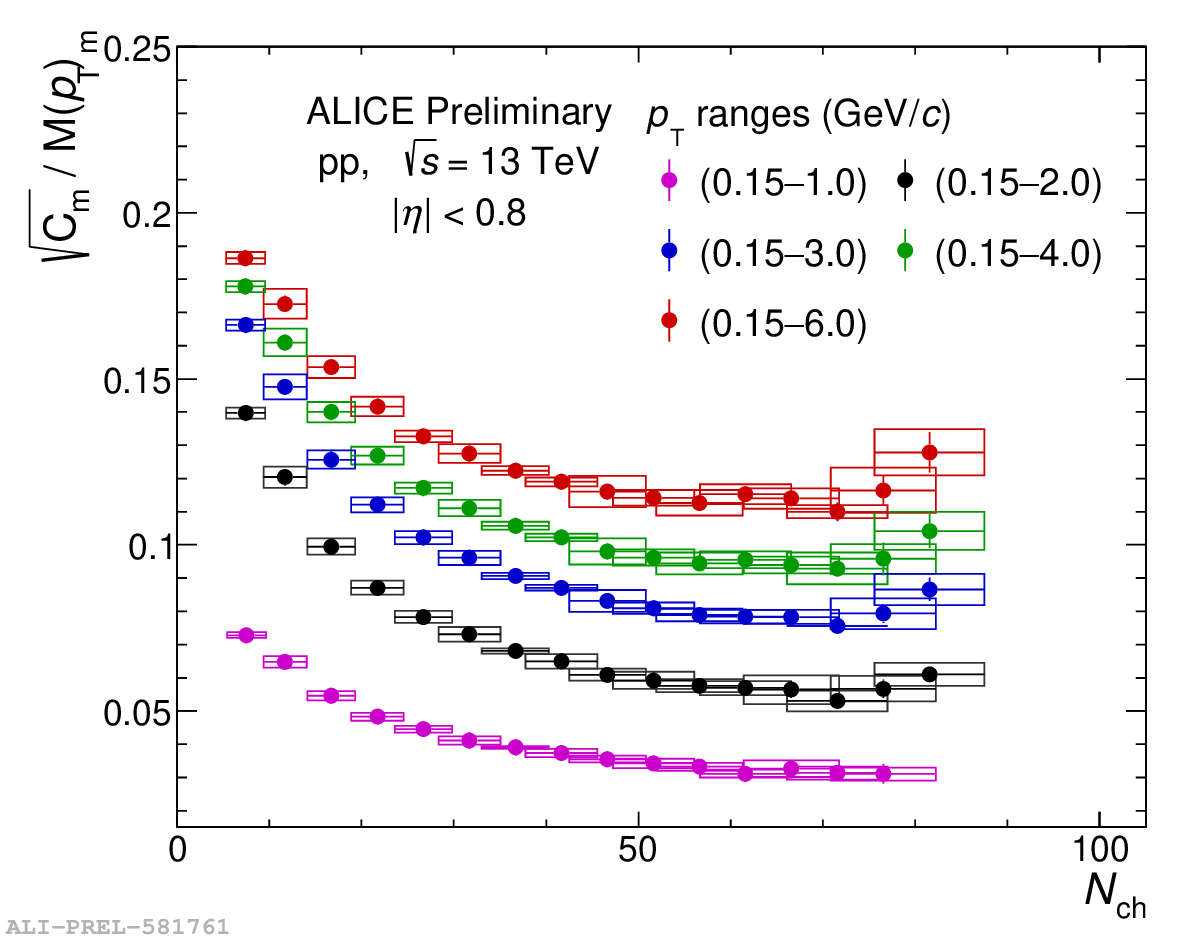}
    \includegraphics[scale=0.36]{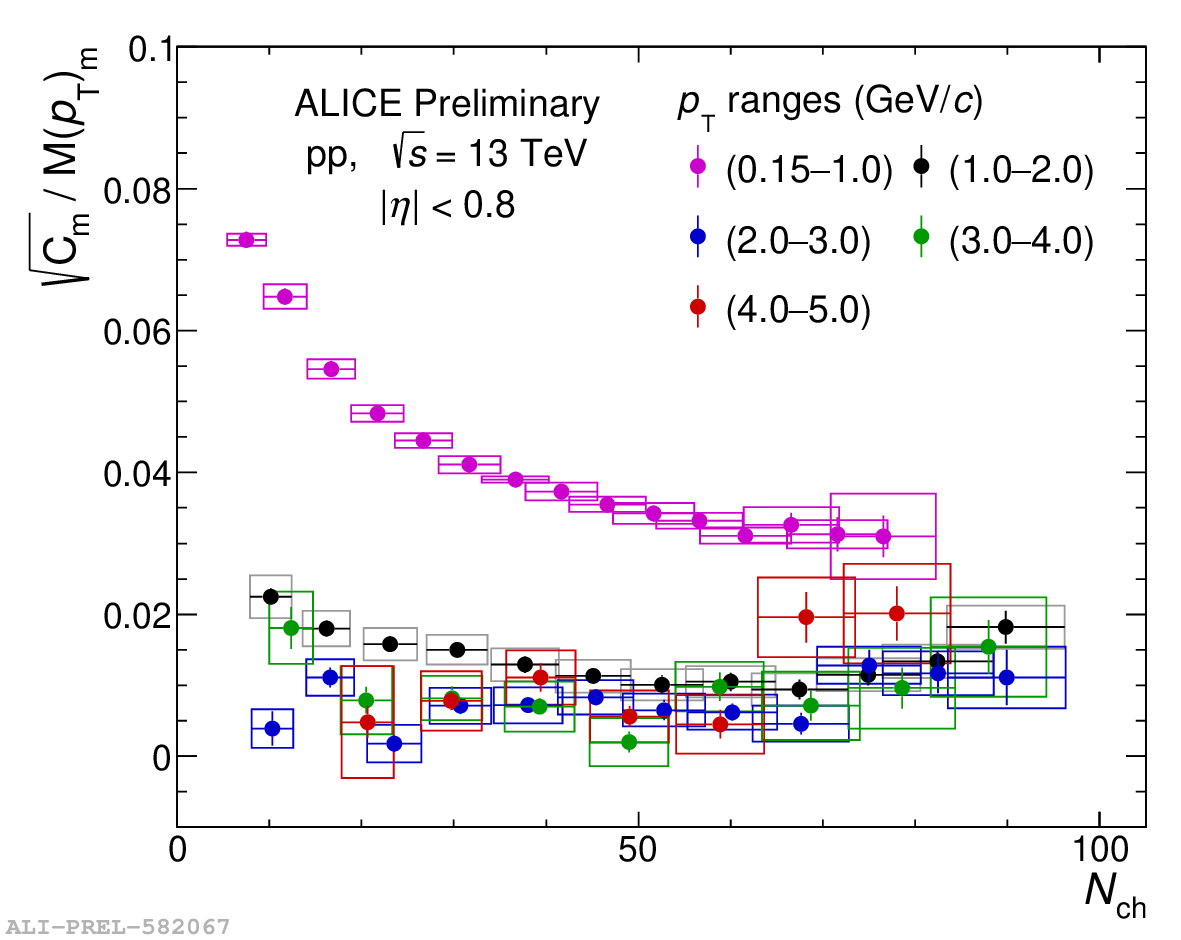}
    \caption{Variations of $\sqrt{C_m}/M(p_{\rm T})_{m}$ with $N_{\rm ch}$ for varying $p_{\rm T}$ window widths on the left and varying $p_{\rm T}$ window positions on the right.}\label{fig1}
\end{figure}
\indent Figure~\ref{fig1} shows the variation of the two-particle correlator, $\sqrt{C_m}/M(p_{\rm T})_{m}$, as a function of charged-particle multiplicity for two different $p_{\rm T}$ selection schemes. The left panel illustrates the effect of progressively widening the $p_{\rm T}$ interval, while the right panel explores the correlator behavior as a function of $p_{\rm T}$ window positions, with fixed width. \\

\indent In the case of widening $p_{\rm T}$ windows, the strength of the correlator increases with multiplicity. This rise is attributed to the inclusion of a larger number of particles within the acceptance, enhancing the probability of correlated pairs contributing to the fluctuation signal. Conversely, when the $p_{\rm T}$ window is shifted across momentum space while keeping its width fixed, a distinct trend emerges. A significant drop in the correlation strength is observed on moving from the lowest $p_{\rm T}$ window (0.15–1.0 GeV/$c$) to a window of higher $p_{\rm T}$ range (1.0–2.0 GeV/$c$). This, in turn, suggests the decreasing influence of soft processes such as collective flow and resonance decays on shifting the $p_{\rm T}$ window to a higher $p_{\rm T}$ range~\cite{23a, 23b}. On shifting the window to still higher $p_{\rm T}$ ranges, the correlation strength is observed to be suppressed further and shows minimal multiplicity dependence. In the highest $p_{\rm T}$ interval, the correlator values approach zero, suggesting a diminishing role of collectivity or correlated particle production mechanisms and an increasing dominance of independent high-momentum transfer processes such as jet fragmentation and hard scatterings~\cite{5, 6, 24}. \\

\begin{figure}[htbp!]
  \centering
    \includegraphics[scale=0.36]{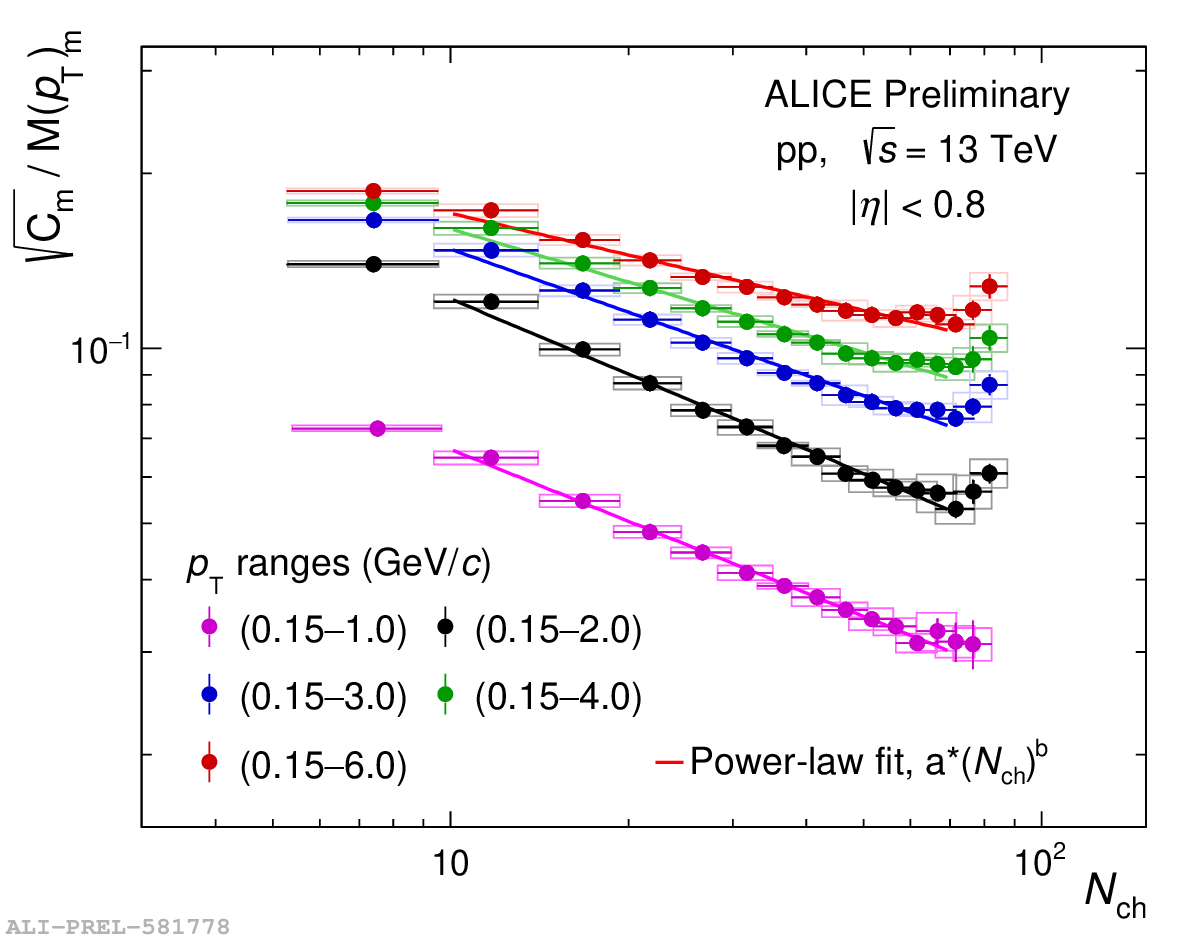}
    \includegraphics[scale=0.36]{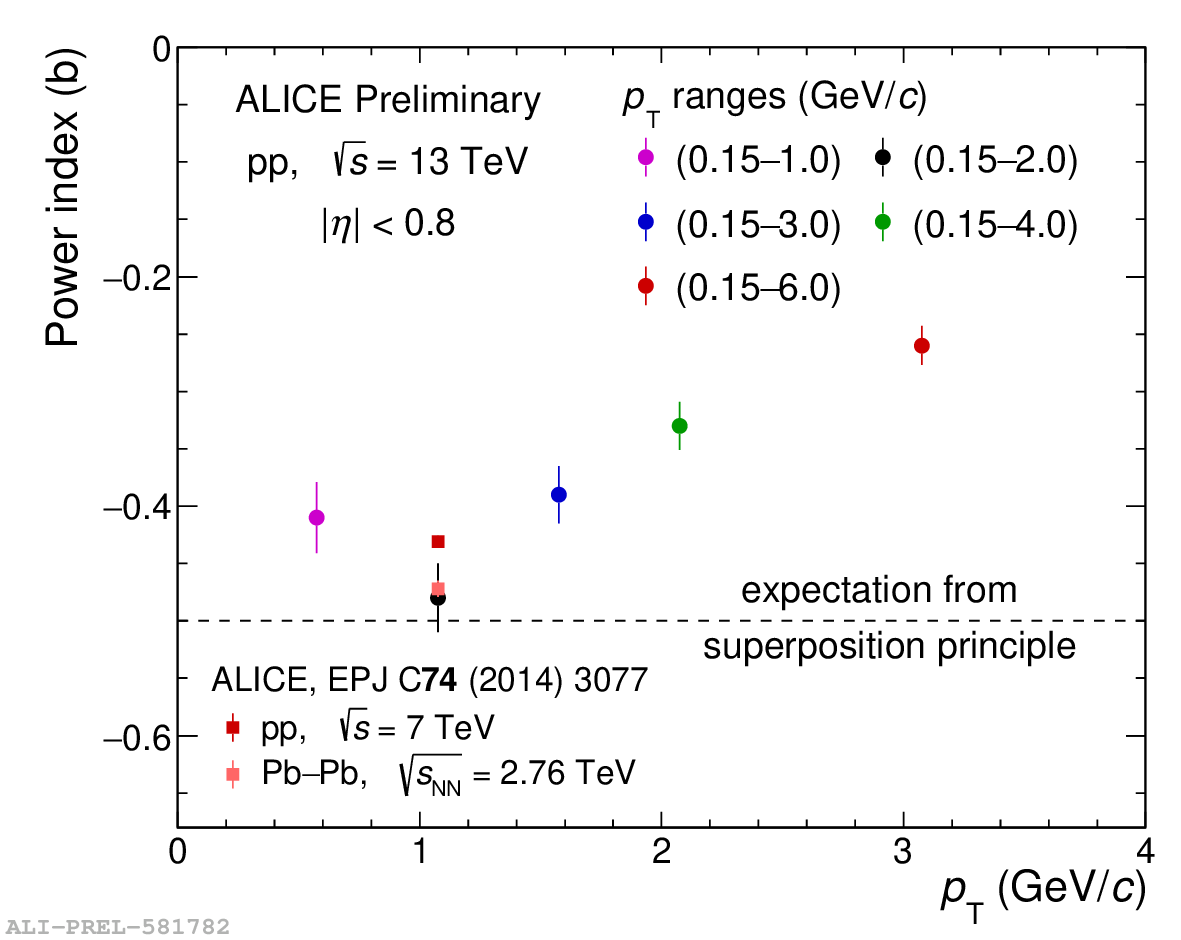}
    \caption{A power-law fit to the normalized correlator for progressively widening $p_{\rm T}$ windows is shown on the left, while a comparison of the power-law index with the expectation from a simple superposition scenario is presented on the right.}\label{fig2}
\end{figure}
\indent Power-law fits to the correlator values against $N_{\rm ch}$ is performed and shown in Fig.~\ref{fig2} for progressively widening $p_{\rm T}$ windows. The fitting function used is $a \cdot (N_{\rm ch})^b$, where $a$ and $b$ are free parameters. The fits are performed in the range $10 < N_{\rm ch} < 70$. In the absence of dynamical correlations, the correlator is expected to follow statistical scaling, yielding a power index of $b = -0.5$. Any deviation from this value would suggest the presence of dynamical correlations. The values of power index $b$ obtained from the best fit to the data are plotted as a function of the $p_{\rm T}$ window widths in the right panel of Fig.~\ref{fig2}. A non-monotonic behavior of $b$ with $p_{\rm T}$ window widths is observed. For the lowest interval (0.15–1.0 GeV/$c$), $b \approx -0.41 \pm 0.031$, indicating a moderate deviation from statistical expectations. With the inclusion of higher $p_{\rm T}$ particles (e.g., up to 2.0 GeV/$c$), $b \approx -0.48 \pm 0.03$, suggesting a significant decrease with increasing multiplicity. However, on further widening of the $p_{\rm T}$ window, an increase in the power index may be noticed in the figure, indicating a reduced sensitivity of the parameter to multiplicity enhancement due to widening of $p_{\rm T}$ window. These observations suggest a crossover between soft and hard contributions to the observed fluctuations. The figure also shows the ALICE measurements~\cite{5} previously reported for pp collisions at $\sqrt{s} = 13$ TeV and Pb–Pb collisions at $\sqrt{s_{\rm NN}} = 2.76$ in the $p_{\rm T}$ range, 0.15-2.0 GeV/$c$. For pp collisions, it is seen that the magnitude of $b$ increases by $\sim 11\%$ from beam energy $\sqrt{s} = 7$ to $13$ TeV.\\

\begin{figure}[htbp!]
 \centering
    \includegraphics[scale=0.82]{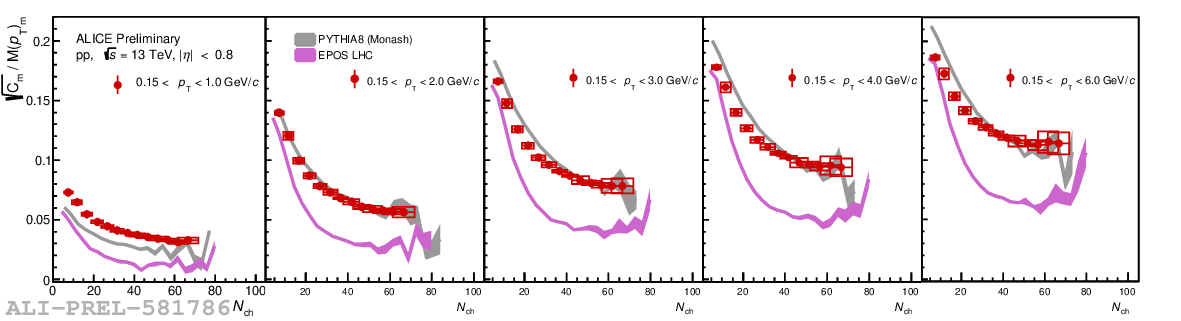}
    \caption{Experimentally observed normalized correlator values against $N_{\rm ch}$ compared with the predictions of PYTHIA and EPOS LHC models for $p_{\rm T}$ windows of increasing widths.}\label{fig3}
    \includegraphics[scale=0.82]{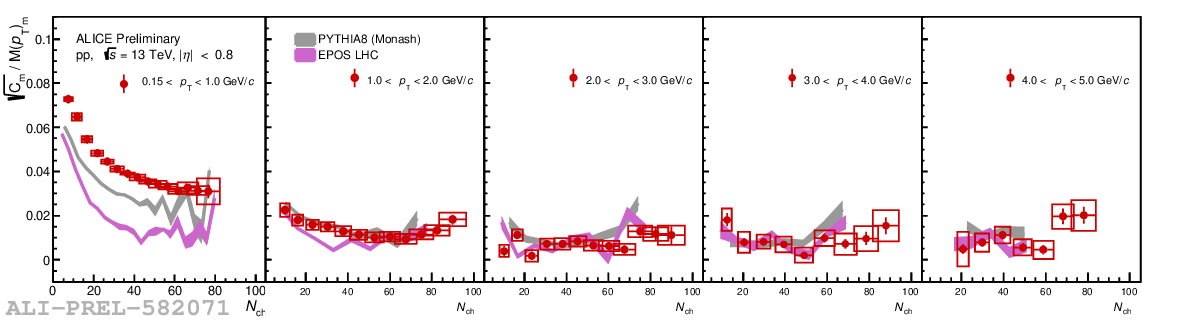}
    \caption{Comparison of the experimentally observed normalized correlator values against multiplicity with PYTHIA and EPOS LHC  predictions for different positions of $p_{\rm T}$ window.}\label{fig4}
\end{figure}

\indent In order to compare the findings based on the real data with the monte carlo (MC) models, PYTHIA8 (Monash tune) and EPOS LHC, a parallel analysis of the MC data has been carried out and the results are displayed in Figs.~\ref{fig3} and~\ref{fig4}. It is interesting to note from the figure that the decreasing trend of the correlator with increasing $N_{\rm ch}$ exhibited by the data is in qualitative agreement with the PYTHIA and EPOS LHC predictions. The models, however, fail to describe the data quantitatively in the low $p_{\rm T}$ range, i.e., 0.15-1.0 GeV/$c$. On increasing the $p_{\rm T}$ window width, which includes higher $p_{\rm T}$ particles, agreement between the data and PYTHIA model may be noticed to be rather better. Such improvements may be attributed to its pQCD-inspired framework, which effectively takes into account the semi-hard and hard processes, including multiple parton interactions and jet fragmentation~\cite{25}. EPOS LHC~\cite{27}, which incorporates the collective effects through a core-corona approach and a hydrodynamic phase, may be noticed to underestimate the correlation strength.\\

\indent Shown in Fig.~\ref{fig4} are the correlator dependence on $N_{\rm ch}$ for fixed-width $p_{\rm T}$ intervals at increasing momentum scales. A sharp drop in correlation strength is observed on moving from the low-$p_{\rm T}$ (0.15–1.0 GeV/$c$) to the intermediate-$p_{\rm T}$ region (1.0–2.0 GeV/$c$), indicating a shift from soft to semi-hard processes. At even higher $p_{\rm T}$, the correlation strength flattens and approaches zero, reflecting the reduced influence of collective dynamics and the growing contribution of uncorrelated hard scatterings~\cite{23a, 23b}. 
\section{Conclusion}
The event-by-event mean $p_{\rm T}$ fluctuations in pp collisions at $\sqrt{s} = 13$ TeV, measured using the ALICE detector, have been examined by varying width and positions of the $p_{\rm T}$ selection windows. In one approach, the $p_{\rm T}$ window is progressively widened to include higher-$p_{\rm T}$ particles. Here, the correlation strength increases with window size, suggesting that the inclusion of particles from semi-hard and hard processes enhances the mean $p_{\rm T}$ correlations. In the second approach, the width of the $p_{\rm T}$ interval is fixed while the window is gradually shifted toward higher $p_{\rm T}$ regions. In this case, the correlation strength decreases as the window moves away from the soft $p_{\rm T}$ region, indicating a transition from soft processes to more uncorrelated semi-hard and hard scatterings that dominate at higher $p_{\rm T}$.\\

\indent Furthermore, the comparison with PYTHIA8 predictions shows that the agreement with data becomes better as more high-$p_{\rm T}$ particles are included. This aligns with the model’s pQCD-based description of hard scattering and parton fragmentation. These observations, thus, tend to suggest that the mean $p_{\rm T}$ correlation strength in pp collisions is mainly driven by semi-hard and hard QCD processes, especially in the high-$p_{\rm T}$ regime.

\end{document}